\documentclass[aps,amsfonts,amsmath,prd,preprint,nofootinbib]{revtex4}
\usepackage{epsf}

\newcommand{\beq}{\begin{equation}}
\newcommand{\eeq}{\end{equation}}

\begin{document}

\title{Probabilities in the landscape}

\author{Alexander Vilenkin}

\address{ Institute of Cosmology, Department of Physics and Astronomy,\\ 
Tufts University, Medford, MA 02155, USA}

\begin{abstract}

I review recent progress in defining probability distributions in the
inflationary multiverse. 

\end{abstract}

\maketitle

\section{Introduction}

It has long been a dream of particle physicists to derive the values
of all constants of Nature from a fundamental theory. With the
development of string theory in the last few decades, it seemed for a
while that we were getting closer to that goal.  String theory is our
best candidate for the fundamental theory, and there has been great
enthusiasm and hope that it will yield a unique set of constants. This
hope, however, appears to have been dashed.  It now appears that
string theory has a multitude of solutions describing vacua with
different values of the low-energy constants. The number of vacua in
this vast ``landscape'' of possibilities can be as large as $10^{500}$
\cite{BP,Susskind,Douglas}.

In the cosmological context, high-energy vacua will drive exponential
inflationary expansion of the universe. Transitions between different
vacua will occur through quantum tunneling, with bubbles of different
vacua nucleating and expanding in the never-ending process of eternal
inflation. As a result, the entire landscape of vacua will be
explored.

If indeed this kind of picture describes our universe, then we will
never be able to calculate all constants of Nature from first
principles. At best we may only be able to make statistical
predictions. The key problem is then to calculate the probability
distribution for the constants. It is often referred to as {\it the
measure problem}.

The probability $P_j$ of observing vacuum $j$ can be expressed as a
product \cite{AV95}
\beq
P_j=P_j^{(V)}n_j^{(obs)},
\label{Pj}
\eeq
where $P_j^{(V)}$ is the fraction of volume occupied by vacuum of type
$j$ and $n_j^{(obs)}$ is the number of observers per unit volume. The
distribution (\ref{Pj}) then gives the probability for a randomly
picked observer to be in a vacuum of type $j$.

The density of observers $n_j^{(obs)}$ cannot at present be
calculated, but in many interesting cases it seems reasonable to
approximate it as being proportional to the density of suitable stars,
which is in turn proportional to the fraction of matter clustered in
giant galaxies \cite{AV96,Weinberg96,MSW}. 

The volume factor $P_j^{(V)}$ presents a problem of a very different
kind: the result depends very sensitively on the choice of a spacelike
hypersurface (a constant-time surface) on which the distribution is
to be evaluated.  This problem was uncovered by Andrei Linde and his
collaborators when they first attempted to calculate volume
distributions \cite{LLM,LM,GBL}. It eluded resolution for more than a
decade, but recently there have been some promising developments,
and I believe we are getting close to completely solving the
problem. The purpose of this paper is to review the new proposals for
$P_j^{(V)}$.

\section{Problem with global-time measure}

The spacetime structure of an eternally inflating universe is
schematically illustrated in Fig.1.  The bubbles expand rapidly
approaching the speed of light, so their worldsheets are well
approximated by light cones. If the vacuum inside a bubble has
positive energy density, it becomes a site of further bubble
nucleation; we call such vacua ``recyclable''.  Negative-energy vacua,
on the other hand, quickly develop curvature singularities; we shall
call them ``terminal vacua''.

\begin{figure}
\begin{center}
\leavevmode\epsfxsize=5in\epsfbox{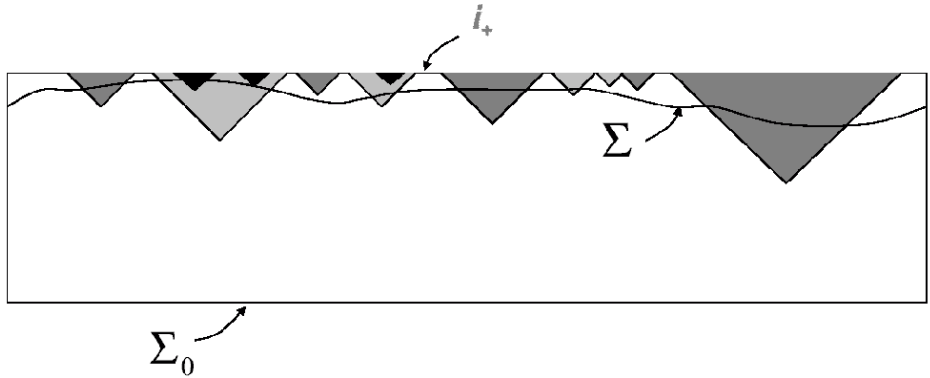}
\caption{A schematic conformal diagram for a comoving region in an 
eternally inflating universe. Bubbles of different vacua are represented 
by different shades of gray. The upper boundary of the diagram $i_+$ is the 
future timelike infinity. A surface of constant global time $\Sigma$ cuts 
through the entire region and intersects many bubbles.} 
\end{center}
\end{figure}

The diagram represents a comoving region, which is initially
comparable to the horizon. The initial moment is a spacelike
hypersurface $\Sigma_0$, represented by the lower horizontal boundary
of the diagram, while the upper boundary represents future infinity,
when the region and all the bubbles become infinitely large.
How can we find the fraction of volume occupied by different vacua? A
natural thing to do is to consider a spacelike hypersurface $\Sigma$,
which cuts through the entire region, as shown in the figure. If $t$
is a globally defined time coordinate, then all surfaces $t={\rm
const}$ will have this property. One can use, for example, the proper
time along the ``comoving'' geodesics orthogonal to the surface
$\Sigma_0$.\footnote{The term ``comoving'' is used very loosely here,
since the vacuum does not define any rest frame. Any congruence of
geodesics orthogonal to a more or less flat spacelike surface
$\Sigma_0$ can be regarded as ``comoving''.}  Alternatively, one could
use the so-called scale factor time, defined as a logarithm of the
expansion factor along the comoving geodesics, or any other suitable
time coordinate. Once the time coordinate is specified, one can find
the fraction of volume occupied by different vacua on the surface
$t={\rm const}$ and then take the limit $t\to\infty$.

Unfortunately, as I have already mentioned, the result of this
calculation is sensitively dependent on one's choice of the time
coordinate \cite{LLM}. The reason is that the volume of an eternally
inflating universe is growing exponentially with time. The volumes of
regions filled with all possible vacua are growing exponentially as
well. At any time, a substantial part of the total volume is in new
bubbles which have just nucleated. Which of these bubbles are cut by
the surface depends on how the surface is drawn; hence the
gauge-dependence of the result. Since time is an arbitrary label in
General Relativity, none of the possible choices of the global time
coordinate appears to be preferred.  For more discussion of this
gauge-dependence problem, see \cite{Guth,Tegmark,Winitzki1}.

\section{A pocket-based measure}

It is now becoming increasingly clear that the solution to the problem
lies in the direction of using the local definition of time within
individual bubbles, or, as Alan Guth called them, ``pocket
universes''. I will first discuss how this works in the simplest case,
when there is a single type of bubble; the general situation will be
considered in the next section.

Suppose we have an eternally inflating universe filled with a
metastable false vacuum $F$, which decays to the true vacuum $T$
through bubble nucleation. The vacuum energy is thermalized within the
bubbles, and in due course observers evolve there and measure the
constants of Nature.  Suppose further that there is a scalar field
$X$, which affects the values of some constants and has a very
slowly varying potential $U(X)$. The values of $X$ are randomized by
quantum fluctuations during inflation, so $X$ is slowly varying in
space within each bubble. We shall assume that the slope of $U(X)$ is
so small that time variation of $X$ is negligible during the epoch
when the observers are present. Our goal is to find the distribution
\beq
P^{(V)}(X)dX, 
\label{PX}
\eeq
which gives the volume fraction occupied by regions where the field is
in the interval between $X$ and $X+dX$.

A measure based on a global time coordinate runs into the same problem
as in the case of a discrete distribution $P_j^{(V)}$: the result for
$P^{(V)}(X)$ is gauge-dependent. There is, however, a simple way
around this difficulty \cite{GTV,AV98,VVW}. It is well known \cite{CdL}
that bubble interiors appear to their inhabitants as self-contained
infinite universes of negative curvature. A natural definition of the
time coordinate in such a universe is to identify it with one of the
physical variables, e.g., the energy density. Each pocket universe
will then have its own set of infinite constant-time surfaces (see
Fig.2).

\begin{figure}
\begin{center}
\leavevmode\epsfxsize=5in\epsfbox{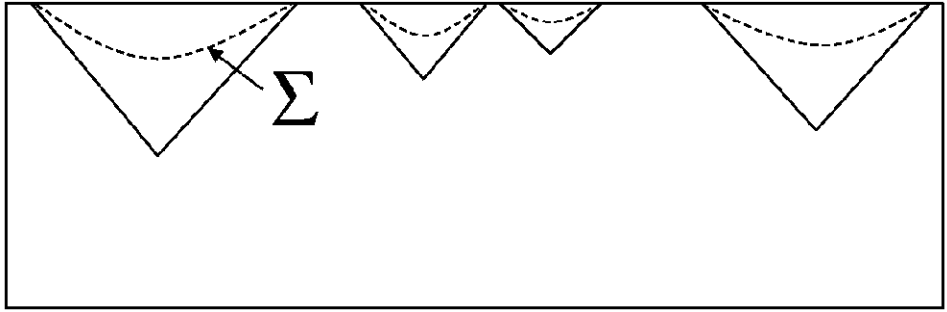}
\caption{A comoving region with a single type of bubble. Bubbles
differ by details of the scalar field distribution, but are
statistically identical. Each pocket is internally an infinite open universe. 
Constant-time surfaces within pockets, shown by dashed lines, 
are infinite spacelike surfaces.} 
\end{center}
\end{figure}

The proposal of \cite{GTV,AV98,VVW} is to calculate the distribution
(\ref{PX}) within a single pocket universe. It does not matter which
one, since all pocket universes are statistically equivalent: they all
have the same volume distribution of $X$.

Once a pocket universe is selected, we choose some constant-time
surface within it as our reference surface $\Sigma$.  The next step is
to find the volume distribution of $X$ on $\Sigma$. Some care is
needed here, since $\Sigma$ is an infinite hypersurface. If one
calculates the distribution in a region of finite size and then takes
the limit of size going to infinity, the result may depend on the
limiting procedure. To avoid this danger, one has to make sure that
the shape of the region is not correlated with the distribution of
$X$. The simplest choice is to use a ``spherical'' region, defined as
a set of points with a geodesic distance $r<R$ from a given center,
with the limit $R\to\infty$ taken afterwards
\cite{AV98,VVW}.\footnote{An alternative definition of the distance is
in terms of the area element, $dA={\tilde r}^2 d\Omega$, with solid
angle $d\Omega$ defining a bunch of radial geodesics. It is possible
that different definitions of $r$ may yield different volume
distributions for $X$, even in the limit $r\to\infty$. The reason is
that in a space of negative curvature, most of the volume of a sphere
is near its surface (${\tilde r}$ is an exponentially growing function
of $r$). The distributions for $X$ may differ if the spacetime
geometry is influenced by the local value of $X$. (This would be
analogous to the gauge-dependence of a global-time measure.) This
problem does not arise if the comoving sphere is defined at an early
time, when the open FRW universe inside the bubble is dominated by
curvature (see Sec.~IV.C).}

Since we assume that $X$ does not change in time, its value remains
constant along the comoving geodesics, and thus the distribution
(\ref{PX}) is time-independent if we think of it as a comoving-volume
distribution.\footnote{The physical-volume distribution will generally
evolve, since the local value of $X$ may affect the expansion rate of
the universe.} The density of observers $n^{(obs)}(X)$ should then
also be understood as the number of observers per unit comoving
volume. If we include all present, past and future observers, then
$n^{(obs)}(X)$ is also time-independent. Moreover, the full
distribution 
\beq
dP(X)\propto P^{(V)}(X)n^{(obs)}(X)
\eeq
does not depend on the choice of the reference hypersurface $\Sigma$,
as long as $P^{(V)}(X)$ and $n^{(obs)}(X)$ are evaluated on the same
hypersurface.

Some analytic and numerical techniques for calculating $P^{(V)}(X)$
have been suggested in \cite{GTV,AV98,VVW}. These techniques are not
yet fully developed, and there are some interesting issues that still
need to be addressed (e.g., the ``ergodic conjecture'' in
Ref.~\cite{GSPVW}).  But as a matter of principle, the problem appears
to have been solved in the case of identical pockets.

\section{Counting pockets}

If there are many different types of pockets, then it is clearly not
sufficient to consider a single bubble. We have to learn how to
compare the numbers of observers in different bubbles. Since the
bubbles are disconnected from one another, we have to define a
comoving length scale $R_j$ on which observers are to be counted in
bubbles of type $j$. In addition, some bubbles may be more abundant
than others, and we have to introduce a frequency factor $p_j$
characterizing the relative abundance of different bubbles. The full
expression for the volume distribution is then given by \cite{markers}
\beq
P_j^{(V)}\propto p_j R_j^3.
\label{PpR}
\eeq
Here I assume for simplicity that there are no continuous fields $X$
that can vary within bubbles. The discussion can be easily extended to
include such variables (see Sec.IV.D). The question now is: How do we
define $p_j$ and $R_j$? We shall consider them in turn.

\subsection{Bubble abundance $p_j$}

The definition of $p_j$ is a tricky business, because the total number
of bubbles is infinite, even in a region of a finite comoving
size.\footnote{The problem of calculating $p_j$ is somewhat similar to
the question of what fraction of all natural numbers are odd. The
answer depends on how the numbers are ordered. With the standard
ordering, $1,2,3,4, ...$, the fraction of odd numbers in a long
stretch of the sequence is 1/2, but if one uses an alternative
ordering $1,3,~2,~5,7,~4, ...$, the result would be 2/3. One could
argue that, in the case of integers, the standard ordering is more
natural, so the correct answer is 1/2. Here we seek an analogous
ordering criterion for the bubbles.} We thus need to introduce some
sort of a cutoff. Here I shall review the procedure recently proposed
in \cite{GSPVW}, which has some attractive properties.

The proposal is very simple: count only bubbles greater than a certain
comoving size $\epsilon$, and then take the limit $\epsilon\to
0$. To define the comoving size, one has to specify a congruence
of ``comoving'' geodesics emanating from some initial spacelike
hypersurface $\Sigma_0$. As they extend to the future, the geodesics
will generally cross a number of bubbles before ending up in one of
the terminal bubbles, where inflation comes to an end. There will also
be a (measure zero) set of geodesics which never hit terminal
bubbles. The starting points of these geodesics on $\Sigma_0$ provide
a mapping of the eternally inflating fractal
\cite{Aryal,Winitzki,Winitzki2}, consisting of points on $i_+$ where
inflation never ends. In the same manner, each bubble encountered by
the geodesics will also be mapped on $\Sigma_0$, and we can define the
``size'' of a bubble as the volume of its image on $\Sigma_0$. (The
volume of a bubble is calculated including all the daughter bubbles
that nucleate within it.)

In an inflating spacetime, geodesics are rapidly diverging, so bubbles
formed at later times have a smaller comoving size. (The comoving size
of a bubble is set by the horizon at the time of bubble nucleation.)
The bubble counting can be done in an arbitrarily small neighborhood
$\delta$ of any point belonging to the ``eternal fractal'' image
on $\Sigma_0$.  Every such neighborhood will contain an infinite
number of bubbles of all kinds and will be dominated by bubbles formed
at very late times and having very small comoving sizes. The resulting
values of $p_j$, obtained in the limit of bubble size $\epsilon\to 0$,
will be the same in all neighborhoods, because of the universal
asymptotic behavior of eternal inflation. 

It is clear that the same result will also hold in any finite-size
region on $\Sigma_0$ (provided that it contains at least one ``eternal
point''), and for any choice of the initial hypersurface $\Sigma_0$.
Moreover, although we use the metric on $\Sigma_0$ to compare the
bubble sizes, the results are unaffected by smooth conformal
transformations of the metric.  Any such transformation will locally
be seen as a linear transformation, which amounts to a constant
rescaling of bubble sizes. In a sufficiently small neighborhood on
$\Sigma_0$, all bubble sizes are rescaled in the same way, so the
bubble counting should not be affected.

The results obtained using this method are also independent of the
initial conditions at the onset of eternal inflation.\footnote{I
assume that any vacuum is accessible through bubble nucleation from
any other vacuum. Alternatively, if the landscape splits into several
disconnected domains which cannot be accessed from one another, each
domain will be characterized by an independent probability
distribution, and our discussion will still be applicable to any of
these domains.}  This is an attractive property, since the initial
conditions are quickly forgotten in an eternally inflating
universe.\footnote{An earlier suggestion of Ref.~\cite{markers} was to
define $p_j$ as the probability for a comoving observer to end up in a
pocket of type $j$. This probability, however, does depend on the
initial state.}

The calculation of bubble abundances, defined in this way, can be
reduced to an eigenvalue problem for a matrix constructed out of the
transition rates between different vacua \cite{GSPVW}.\footnote{The 
calculation in \cite{GSPVW} assumes that the divergence of
geodesics is everywhere determined by the local vacuum energy
density. This is somewhat inaccurate, since it ignores the brief
transition periods following the bubble crossings and the focusing
effect of the domain walls. The accuracy of the method is expected to be
up to factors $O(1)$. A more detailed discussion will be given
elsewhere \cite{walls}.} This
prescription has been tried in some simple models and appears to give
reasonable results \cite{GSPVW,SPV}. For example, if there is a single
false vacuum, which can decay into a number of vacua with nucleation
rates $\Gamma_j$ , one finds
\beq
p_j\propto \Gamma_j,
\eeq
as intuitively expected.

\subsection{An equivalent proposal}

An alternative prescription for $p_j$ has been suggested by Easther,
Lim and Martin \cite{ELM}. They randomly select a large number
$N$ of worldlines out of a congruence of comoving geodesics and define
$p_j$ as being proportional to the number of bubbles of type $j$
intersected by at least one of these worldlines in the limit of
$N\to\infty$.

As the number of worldlines is increased, the average comoving
distance $\epsilon$ between them gets smaller, so most bubbles of
comoving size larger than $\epsilon$ are counted.  In the limit of
$N\to\infty$, we have $\epsilon\to 0$, and it is not difficult to see
that this prescription is equivalent to the one described in the
preceding subsection. (For a rigorous proof, see {\it Note added} in
\cite{GSPVW}.)

\subsection{Reference length $R_j$}

Our next task is to define the comoving reference scale $R_j$. The
first thing that comes to mind is to set $R_j$ to be the same for all
bubbles. However, this is not enough. The expansion rate is different
in different bubbles, so the physical length scales corresponding to
$R_j$ will not stay equal, even if they were equal at some moment. We
could specify the times $t_j$ at which $R_j$ are set to be equal, but
any such choice would be subject to the criticism of being arbitrary.

A possible way around this difficulty was proposed in \cite{GSPVW}. At
early times after nucleation, the dynamics of open FRW universes
inside bubbles is dominated by curvature, with the scale factor given
by
\beq
a_j(t)\approx t
\label{at}
\eeq
for all types of bubbles. For example, for a quasi-de Sitter bubble
interior,
\beq
a_j(t)\approx H_j^{-1}\sinh (H_j t),
\eeq
where $H_j=(8\pi G\rho_j/3)^{1/2}$ is the expansion rate corresponding
to the local vacuum energy density $\rho_j$.  The specific form of the
scale factor at late times is not important for our argument. The
point is that for $t\ll H_j^{-1}$ all bubbles are nearly identical,
with the scale factor (\ref{at}). (This is basically a consequence of
the universal spacetime structure in the vicinity of the light cone $t=0$.)

The proposal of \cite{GSPVW} is that the reference scales should be
chosen so that $R_j$ are the same at some small $t=\tau$. The
choice of $\tau$ is unimportant, as long as $\tau\ll H_j^{-1}$ for all
$j$. Then, up to a constant, the physical length corresponding to
$R_j$ is
\beq
R_j(t)=a_j(t).
\label{Ra}
\eeq
For times $t\gg H_j^{-1}$, this can be expressed as 
\beq
R_j(t)\approx H_j^{-1}Z_j(t),
\label{RZ}
\eeq
where $Z_j$ is the expansion factor since the onset of the
inflationary expansion inside the bubble ($t\sim H_j^{-1}$).

Alternatively, $R_j$ in (\ref{RZ}) can be identified as the curvature
scale. It is the characteristic large-scale curvature radius of the
bubble universe. This definition makes no reference to early times
close to the bubble nucleation: the curvature radius can be found at
any time. It is, in principle, a measurable quantity.

\subsection{Continuous variables}

The prescription (\ref{PpR}) can be straightforwardly generalized to
the case when, in addition to bubbles, there are some continuously
varying fields $X$:
\beq
P_j^{(V)}(X)\propto p_j {\hat P}_j(X)R_j^3(X).
\label{hatP}
\eeq
Here, ${\hat P}_j(X)$ is the normalized distribution for $X$ in a
bubble of type $j$ at $t=\tau \ll H_j^{-1}$,
\beq
\int{\hat P}_j(X)dX=1.
\eeq
This distribution can be calculated analytically or numerically, using
the methods of Refs.~\cite{VVW,GSPVW}.

\section{Discussion}

The above definition of the measure is just a proposal. We have not
derived it from first principles. In fact, there is no guarantee that
there is some unique measure that can be used for making predictions
in the multiverse. How, then, can we ever know that we made the right
choice out of all possible options?

What I find encouraging is that even a single definition of measure
that satisfies some basic requirements proved very difficult to find.
In addition to being mathematically consistent, we require that the
measure should not depend on any arbitrary choices, such as the choice
of gauge or of a spacelike hypersurface, and that it should be
independent of the initial conditions at the onset of inflation. It
would be interesting to know how much freedom is left by these
requirements. In other words, how uniquely do they specify the
proposal of Ref.~\cite{GSPVW}?

The measure can further be tested by working out simple models and
checking whether or not the resulting distributions are reasonable. So
far, the proposal of \cite{GSPVW} seems to have passed this test. 

A possible alternative, advocated in \cite{Tegmark}, is to assume that
``all infinities are equal'' and set 
\beq
p_j={\rm const}
\label{pconst}
\eeq 
for all $j$. This prescription is clearly gauge-invariant and is
independent of the initial conditions. The resulting measure, however,
appears to be counter-intuitive. Intuitively, one expects that,
everything else being equal, the probability assigned to a certain
type of bubble should be proportional to the bubble nucleation rate.
This is satisfied for the proposal of \cite{GSPVW}, but if
(\ref{pconst}) is adopted, the probability would be completely
independent of the nucleation rate.

The ultimate test of any proposed measure will be a comparison of its
predictions with observations. The first attempts in this direction
have already produced some intriguing results
\cite{Aguirre,Freivogel,Hall1,QLambda,Tegmark2,Hall2,SPV,Hawking}. It seems
safe to predict that we will hear more on this subject in the future.

~~~~~~~~~~~~~~~~~~~

I am grateful to Leonard Susskind for pointing out an error in
the original version of this paper and for stimulating
correspondence. I am also grateful to Jaume Garriga, Delia
Schwartz-Perlov and Serge Winitzki for discussions and useful
comments.  This work was supported in part by the National Science
Foundation.


\begin{thebibliography}{99}

\bibitem{BP}
R.~Bousso and J.~Polchinski, JHEP {\bf 0006}, 006 (2000).

\bibitem{Susskind}
L.~Susskind, ``The anthropic landscape of string theory,''
arXiv:hep-th/0302219.

\bibitem{Douglas}
M.R. Douglas, JHEP {\bf 0305}, 046 (2003).

\bibitem{AV95}
A. Vilenkin, Phys.\ Rev.\ Lett.\ {\bf 74}, 846 (1995).

\bibitem{AV96}
A. Vilenkin, in {\it Cosmological Constant and the Evolution of the
Universe}, ed by K. Sato, T. Suginohara and N. Sugiyama (Universal
Academy Press, Tokyo, 1996).

\bibitem{Weinberg96}
S. Weinberg, in {\it Critical Dialogues in Cosmology}, ed. by
N.~G. Turok (World Scientific, Singapore, 1997).

\bibitem{MSW}
H.~Martel, P.~R.\ Shapiro and S.~Weinberg, Ap.J.\ {\bf 492},
29 (1998).

\bibitem{LLM}
A.~D.~Linde, D.~A.~Linde, and A.~Mezhlumian, Phys.\ Rev.\ D {\bf 49},
1783 (1994)
 
\bibitem{LM}
A.~D.~Linde and A.~Mezhlumian, Phys.\ Rev.\ D {\bf 53}, 4267 (1996)

\bibitem{GBL}
J. Garcia-Bellido and A. D. Linde, Phys.~Rev.~D {\bf 51}, 429 (1995).

\bibitem{Guth}
A.~H.~Guth, Phys.\ Rept.\ {\bf 333}, 555 (2000)

\bibitem{Tegmark}
M.~Tegmark, JCAP {\bf 0504}, 001 (2005)

\bibitem{Winitzki1}
S. Winitzki, Phys. Rev. D 71, 123507 (2005).

\bibitem{GTV}
J.~Garriga, T.~Tanaka, and A.~Vilenkin, Phys.\ Rev.\ D {\bf 60},
023501 (1999)

\bibitem{AV98}
A.~Vilenkin, Phys.\ Rev.\ Lett.\ {\bf 81}, 5501 (1998)

\bibitem{VVW}
V.~Vanchurin, A.~Vilenkin, and S.~Winitzki, Phys.\ Rev.\ D {\bf 61},
083507 (2000)

\bibitem{CdL}
S Coleman and F. DeLuccia, Phys. Rev. {\bf D21}, 3305 (1980).
 
\bibitem{GSPVW}
J. Garriga, D. Schwartz-Perlov, A. Vilenkin and S. Winitzki,
JCAP {\bf 0601}, 017 (2006).

``Probabilities in the inflationary multiverse'',
hep-th/0509184. 

\bibitem{markers}
J.~Garriga and A.~Vilenkin, Phys.\ Rev.\ D {\bf 64}, 023507 (2001)
 
\bibitem{Aryal}
M. Aryal and A. Vilenkin, Phys. Lett. {\bf B199}, 352 (1987).

\bibitem{Winitzki}
S. Winitzki, Phys. Rev. {\bf D65}, 083506 (2002).

\bibitem{Winitzki2} 
S. Winitzki, Phys.  Rev. {\bf D71}, 123523 (2005).

\bibitem{walls}
J. Garriga and A. Vilenkin, unpublished.

\bibitem{SPV}
D. Schwartz-Perlov and A. Vilenkin, ``Probabilities in the
Bousso-Polchinski multiverse'', hep-th/0601162.

\bibitem{ELM} 
R. Easther, E.A. Lim and W.R. Martin, ``Counting pockets
with worldlines in eternal inflation'', astro-ph/0511233.

\bibitem{Aguirre}
A. Aguirre and M. Tegmark, JCAP 0501, 003 (2005).

\bibitem{Freivogel}
B. Freivogel, M. Kleban, M. Rodrigues Martinez, and L. Susskind,
``Observational consequences of a landscape'', hep-th/0505232.

\bibitem{Hall1}
B. Feldstein, L.J. Hall and T. Watari, Phys. Rev. {\bf D72}, 123506
(2005).

\bibitem{QLambda}
J. Garriga and A. Vilenkin, ``Anthropic predictions for $\Lambda$ and
the $Q$ catastrophe'', hep-th/0508005.

\bibitem{Tegmark2}
M. Tegmark, A. Aguirre, M. Rees and F. Wilczek, ``Dimensionless
constants, cosmology and other dark matters'', astro-ph/0511774.

\bibitem{Hall2}
L.J. Hall, T. Watari and T.T. Yanagida, ``Taming the runaway problem
of inflationary landscapes'', hep-th/0601028.

\bibitem{Hawking}
S.W. Hawking and T. Hertog, ``Populating the landscape: a top down
approach'', hep-th/0602091.

\end{thebibliography}
\end{document}